\documentclass[reprint,showpacs,preprintnumbers,superscriptaddress,aps]{revtex4-1}

\usepackage{epsf}
\usepackage{epsfig}
\usepackage{graphicx}
\usepackage{amsmath}
\usepackage{amssymb}

\newcommand{\unit}[1]{\ensuremath{\, \mathrm{#1}}}

\begin{document}

\title{Laser-induced effects on the electronic features of graphene nanoribbons}
\author{Hern\'{a}n L. Calvo}
\affiliation{Institut f\"{u}r Theorie der Statistischen Physik, RWTH
Aachen University, 52056 Aachen, Germany and JARA - Fundamentals of
Future Information Technology.}
\author{Pablo M. P\'{e}rez-Piskunow}
\affiliation{Instituto de F\'{\i}sica Enrique Gaviola (IFEG-CONICET) and
FaMAF, Universidad Nacional de C\'{o}rdoba, Ciudad Universitaria, 5000
C\'{o}rdoba, Argentina.}
\author{Stephan Roche}
\affiliation{CIN2(ICN-CSIC) and Universidad Aut\'{o}noma de Barcelona, 
Catalan Institute of Nanotechnology, Campus UAB, 08193 Bellaterra (Barcelona), Spain.}
\affiliation{ICREA, Instituci\'{o} Catalana de Recerca i Estudis Avan\c{c}ats, 08070 Barcelona, Spain.}
\author{Luis E. F. Foa Torres}
\email{Electronic mail: lfoa@famaf.unc.edu.ar}
\affiliation{Instituto de F\'{\i}sica Enrique Gaviola (IFEG-CONICET) and
FaMAF, Universidad Nacional de C\'{o}rdoba, Ciudad Universitaria, 5000
C\'{o}rdoba, Argentina.}

\begin{abstract}
We study the interplay between lateral confinement and photon-induced
 processes on the electronic properties of illuminated graphene
 nanoribbons. We find that by tuning the device setup (edges geometries,
 ribbon width and polarization direction), a laser with frequency
 $\Omega$ may either not affect the electronic structure, or induce
 bandgaps or depletions at $\pm\hbar\Omega/2$, and/or at other energies
 not commensurate with half the photon energy. Similar features are also
 observed in the dc conductance, suggesting the use of the polarization
 direction to switch on and off the graphene device. Our results could
 guide the design of novel types of optoelectronic nano-devices.
\end{abstract}

\maketitle

The extraordinary properties of graphene \cite{Geim2009, CastroNeto2009,
Dubois2009} led to an unprecedented narrowing in the expected gap
between the understanding of new phenomena and the development of
disruptive applications \cite{Geim2009}. Though originally focused
mainly on pure electronic, mechanical or optical properties, much
attention is now devoted to the interplay between these variables
\cite{Bonaccorso2010}. Graphene optoelectronics \cite{Bonaccorso2010,
Xia2009, Gabor2011, Karch2011, Koppens2011, Ren2009, McIver2012}, in
particular, is one of the most active and promising fields with flagship
applications including energy harvesting devices \cite{Gabor2011} and
novel plasmonic properties \cite{Koppens2011,Chen2012}.

Recently, the captivating possibility of controlling the electronic
properties of graphene through simple illumination with a laser field
\cite{Syzranov2008, Oka2009} has been re-examined through atomistic
calculations \cite{Calvo2011}, calculations of the optical response
\cite{Zhou2011, Busl2012} and proposals for tuning the topological
properties of the underlying photon-induced states \cite{Kitagawa2011,
Gu2011, Dora2012, SuarezMorell2012}, among other interesting issues
\cite{Abergel2009, Kibis2010, Savelev2011, Iurov2012, Liu2012,
San-Jose2012}. The basic idea is that laser illumination may couple
states on each side of the charge neutrality point inducing a bandgap at
$\pm \hbar \Omega/2$, if the field intensity and frequency are
appropriately tuned. This non-adiabatic and non-perturbative effect
relies crucially on the low dimensionality and peculiar electronic
structure of graphene and has attracted much recent attention
\cite{Kitagawa2011, Gu2011, Dora2012,
SuarezMorell2012}. Notwithstanding, most of these predictions were
restricted to bulk graphene. One may wonder about the possible
consequences of reduced dimensionality and quantum confinement.

\begin{figure}[tbp]
\includegraphics[width=8.5cm]{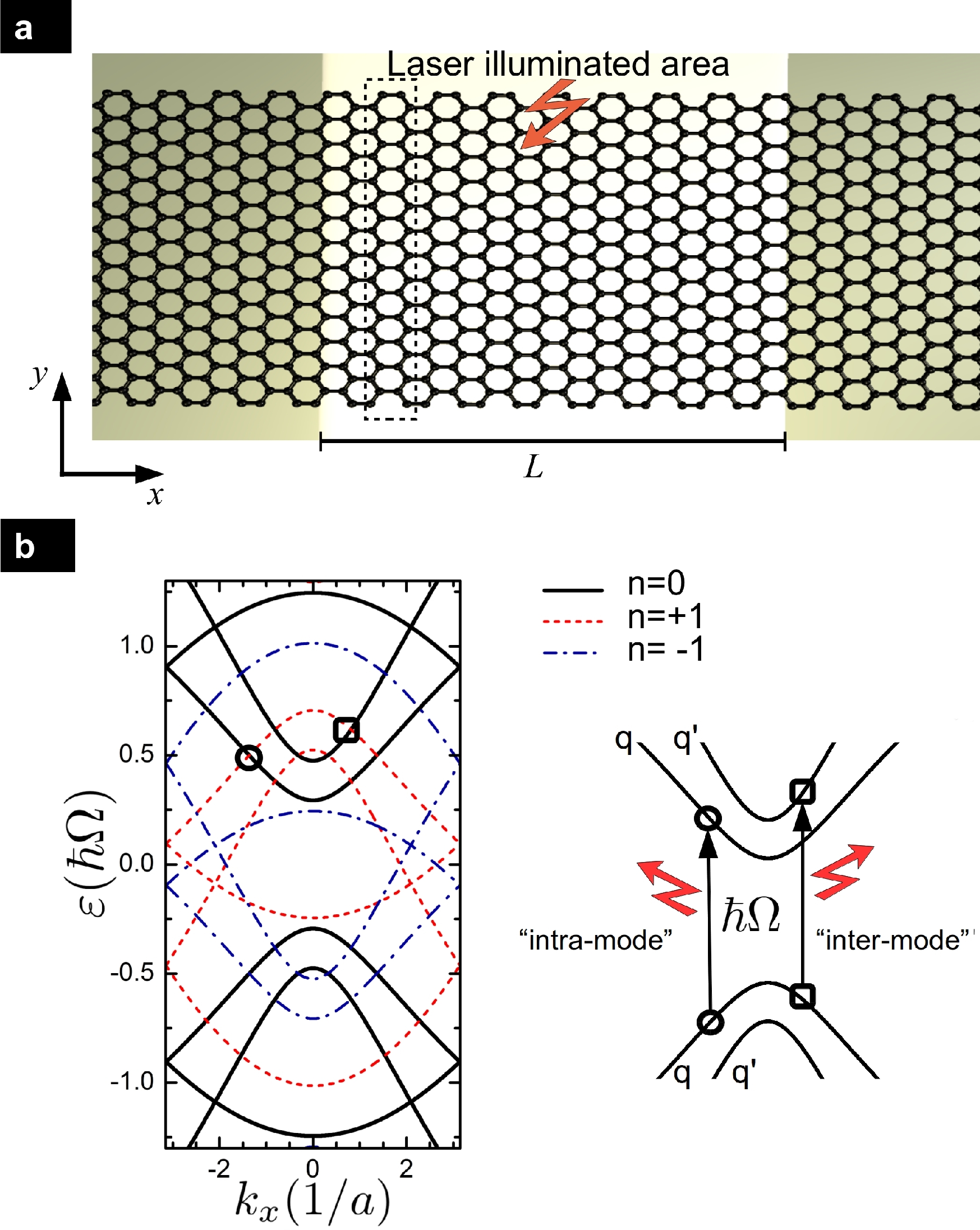}
\caption{(Color online) $a$- Scheme of the considered setup: A
 graphene nanoribbon illuminated by a mid-infrared laser within a finite
 region of length $L$. The unit cell for the case of an armchair ribbon
 is marked with a rectangle (dashed line) and contains $4N$ atoms in the
 notation used in the text. 
$b$- Floquet quasi-energy dispersion for a small ribbon ($N=2$). For the sake of better
 visualization, we show the quasi-energy spectrum without radiation
 (this allows to better distinguish the crossings between levels) and
 choose arbitrary frequency and intensity values. More realistic results
 are shown in Figs.~\ref{fig2} and \ref{fig3}. The scheme on the right
 shows the electronic states for two of such transitions.} 
\label{fig1}
\end{figure}

Here, we address the effects of laser illumination on graphene
nanoribbons and show that lateral confinement plays a crucial role:
tuning the sample size and the direction between the laser polarization
relative to the sample edges, linearly polarized light may or not induce
bandgaps or depletions in the density of states and the conductance
spectra. Strikingly, for finite size samples these features may appear
at energies different from integer multiples of $\pm \hbar
\Omega/2$. This is in stark contrast with bulk graphene where the
electrical response is insensitive to the linear polarization
direction. Our results fill the gap in the understanding of the
laser-induced control of the electrical response and may guide the
design of new experiments on optoelectronic devices.

\textit{Hamiltonian model and solution scheme.} We consider an infinite
graphene nanoribbon illuminated by a laser only in a finite region of
length $L$ and perpendicular to it (as shown schematically in
Fig.~\ref{fig1}-$a$). Electrons in the graphene ribbon are modeled through
a nearest neighbours $\pi$-orbitals Hamiltonian
\cite{Dubois2009,SaitoBook}:
$H_{e}=\sum_{i}E_{i}^{{}}c_{i}% 
^{+}c_{i}^{{}}-\sum_{\left\langle i,j\right\rangle }[\gamma_{i,j}c_{i}%
^{+}c_{j}^{{}}+\mathrm{H.c.}]$, where $c_{i}^{+}$ and $c_{i}^{{}}$
are the electronic creation and annihilation operators at site
$i$, $E_{i}$ are the on-site energies and $\gamma_{i,j}$ the
nearest-neighbors carbon-carbon hoppings amplitudes which are taken
equal to $\gamma_0=2.7 \unit{eV}$ \cite{Dubois2009}. Radiation is
described through a time-dependent electric field $\mathbf{E}$. By
choosing a gauge such that $\mathbf{E}=-\partial \mathbf{A}/ \partial
t$, where $\mathbf{A}$ is the vector potential, the hopping matrix
elements acquire a time-dependent phase according to: $\gamma_{ij}=
\gamma_{0}\exp\left(\mathrm{i}\frac{2\pi}{\Phi_0}\int_{
\mathbf{r}_i}^{\mathbf{r}_j}\mathbf{A}(t)\cdot\mathrm{d}\mathbf{r}\right)$,
where $\Phi_0$ is the magnetic flux quantum. 

Retaining \textit{non-perturbative} and \textit{non-adiabatic}
corrections to the electrical response is crucial for the results
presented hereafter. In this regime, Floquet theory provides an
appropriate framework. An efficient solution using this
scheme is used to obtain the average density of states and the dc
component of the conductance, which is computed from the inelastic
transmission probabilities in Floquet space \cite{Kohler2005}. The
interested reader may find further generalities of the method in
Refs.~\cite{Kohler2005, FoaTorres2005}, while more technical details
will be published elsewhere \cite{JPCM2012}. For a periodic modulation
of the hoppings, the spectral and transport properties can be derived
from the so-called Floquet Hamiltonian: $H_F= H_e-i\hbar
\partial/\partial t$. Such Hamiltonian has a time-independent
representation in the Floquet space, which is the direct product between
the usual Hilbert space and the space of time-periodic functions with
the same period as the Hamiltonian $H_e$
\cite{Shirley,Kohler2005}. Therefore, on top of the $k$ label, our
states have a second label which indicates the number $n$ of photon
excitations in the system. In the absence of radiation, the
quasi-energies spectrum of $H_F$ are given by $\varepsilon(\mathbf{k},
n)=\varepsilon_0(\mathbf{k}) + n\hbar\Omega$
($\varepsilon_0(\mathbf{k})$ is the spectrum of $H_e$).

\textit{Electronic properties of irradiated graphene nanoribbons.} The
first question we address is whereas the response of graphene
nanoribbons to a laser is sensitive to the (linear) polarization
direction. While in the bulk limit both the conductance and the density
of states are independent on the polarization direction, the picture
turns out to change radically in confined geometries.

\begin{figure}[tbp]
\includegraphics[width=8.5cm]{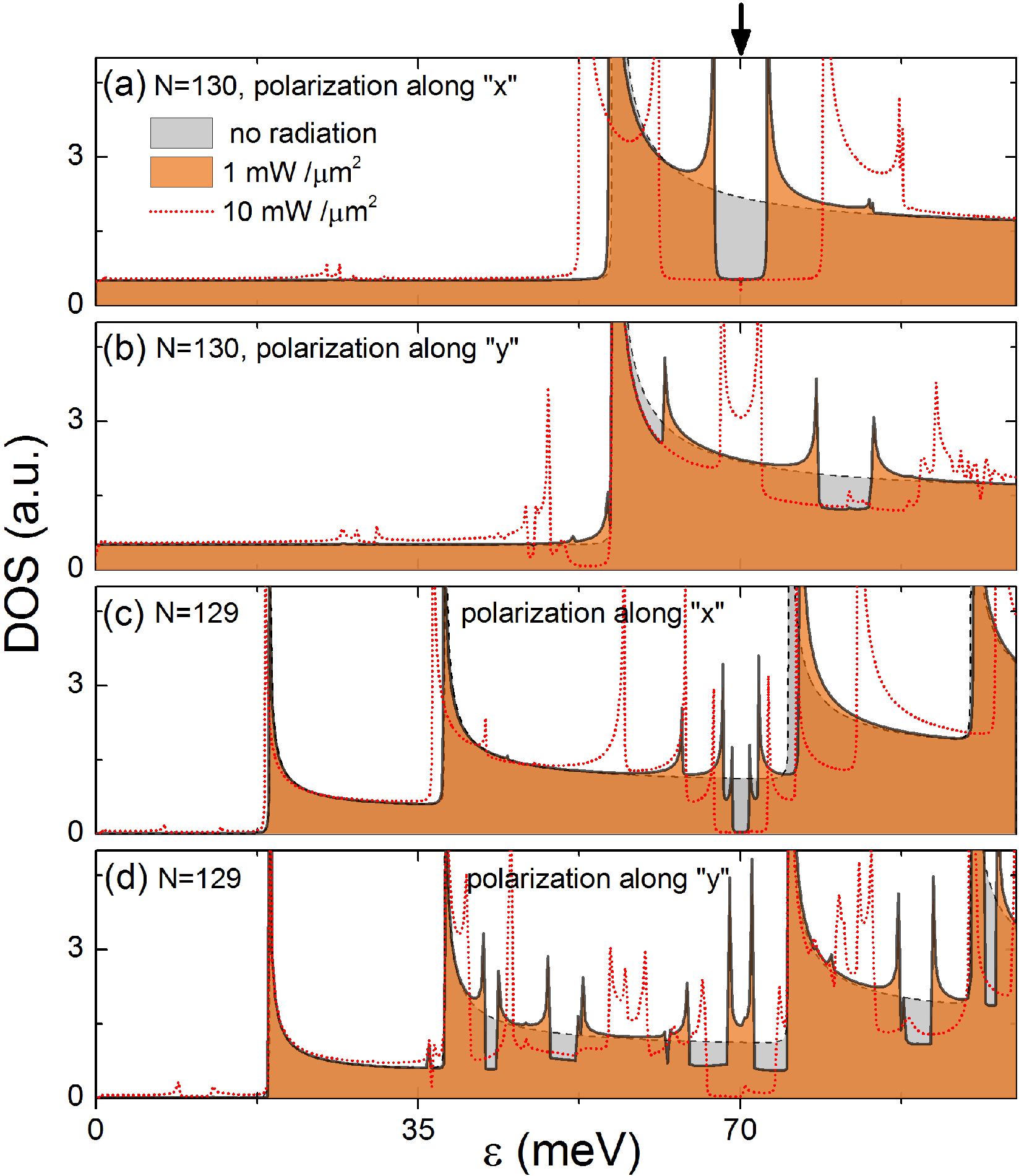}
\caption{(Color online) Average density of states for armchair graphene
 nanoribbons with $N=130$ ($a$,$b$) and $N=129$ ($c$,$d$) for two values
 of the laser power $P$: $1\unit{mW/\mu m^2}$ (solid black line with
 orange shaded area) and $10\unit{mW/\mu m^2}$ (dotted red line). Panels
 $a$ and $c$ ($b$ and $d$) are for polarization along the $x$ direction
 ($y$ direction). The energy corresponding to $\hbar\Omega/2$ is marked
 with an arrow for reference in the top of the panel. The DOS in absence
 of radiation is also shown for comparison (dashed lines, grey shaded
 area).} 
\label{fig2}
\end{figure}

\begin{figure}[tbp]
\includegraphics[width=8.5cm]{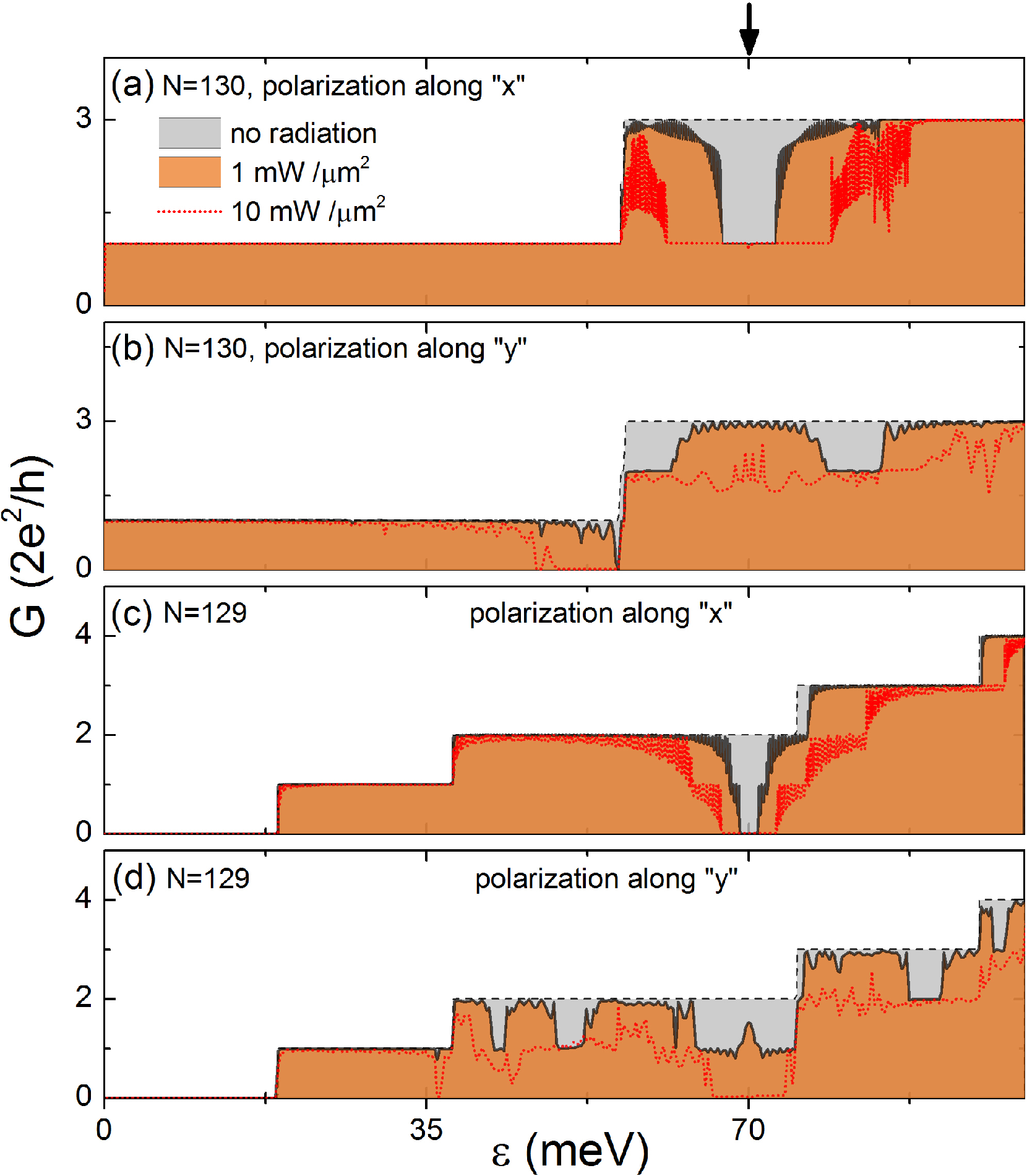}
\caption{(Color online) DC conductance for armchair graphene nanoribbons
 with $N=130$ ($a$,$b$) and $N=129$ ($c$,$d$). Panels $a$ and $c$ ($b$
 and $d$) are for polarization along the $x$ direction ($y$ direction).} 
\label{fig3}
\end{figure}

Figure \ref{fig2} shows the average density of states as a function of
the Fermi energy for a frequency in the mid-infrared regime ($ \hbar
\Omega=140\unit{meV}$). Two different armchair nanoribbons sizes
and polarizations are chosen: $N=130$ ($a$ and $b$), $N=129$ ($c$ and
$d$); and linear polarization along the $x$ ($a$ and $c$) and $y$
($b$ and $d$) directions. For $N=130$ one sees the appearance of strong
depletions at $\pm \hbar \Omega/2$. These depletions are located at the
same energy as the ones for the bulk system \cite{Calvo2011} and
correspond to the excitation of an electron between the conjugate states
at $\pm \hbar \Omega/2$. In striking contrast, the DOS is restored at
$\pm \hbar \Omega/2$ for y-polarized laser, whereas new features occur
at energies incommensurate with $\hbar \Omega$. 

The DOS for the ribbon with $N=129$ ($c$ and $d$) also exhibit a
laser-induced fragmentation of the spectrum, with in this case the
observation of fully depleted energy regions (bandgap) at $\pm \hbar
\Omega/2$. One observes that by increasing the laser intensity, some DOS
modifications are further enhanced (see for instance the DOS depletion
around $\hbar \Omega/2$ for Fig.~\ref{fig2}-$a$), complemented by the
emergence of new fine structure. Figure \ref{fig3} shows the dc
conductance as a function of the Fermi energy position for the same
cases shown in Fig.~\ref{fig2}. Here, we see that the depletions in the
DOS have the same conductance fingerprints. One can see that
\textit{switching the polarization direction may produce a marked on-off
ratio if the Fermi energy is appropriately tuned}.

To rationalize these differences, it is instructive to write $H_e$ in a
basis of independent transversal channels or modes as discussed in
Refs.~\cite{Rocha2010,Zhao2009}. In the absence of radiation, the
Floquet spectrum of the system is just the sum of the contributions from
each of the modes $\varepsilon_0 (\mathbf{k})$ plus their Floquet
replicas: $\varepsilon(\mathbf{k},n)=\varepsilon_0 (\mathbf{k})+n\hbar
\Omega$. An illustration for a very small system is shown in
Fig.~\ref{fig1}-$b$, each mode contains an electron and a hole branch. At
the crossings between the different lines, one may expect for the
effects of radiation to be stronger (if it provides the necessary
coupling between the corresponding states).

From the Floquet spectrum in Fig.~\ref{fig1}-$b$ one can see that there
are two kind of crossings: the ones that connect an electronic (hole)
state with a hole (electron) state belonging to the same mode plus or
minus an integer number of photons (like the one marked with an open
circle); and the ones that connect states in different modes (as the one
marked with an open square in Fig.~\ref{fig1}-$b$). In such cases, a
non-vanishing matrix element of the Floquet Hamiltonian will give intra
and inter-mode transitions respectively. Given the electron-hole
symmetry of the spectrum, intra-mode transitions always connect states
which are symmetric relative to the Dirac point, i.e. integer multiples
of $\pm \hbar \Omega/2$. On the other hand, inter-mode transitions
always couple states which are not symmetrically located from the Dirac
point, as can be seen on Fig.~\ref{fig1}-$b$. A scheme showing these two
type of transitions is shown in Fig.~\ref{fig1}-$b$.

For armchair graphene nanoribbons, it turns out that a laser with linear
polarization along the transport direction ($x$) does not mix these
transversal modes, leading to features in the density of states
\textit{only} at $\pm n \hbar \Omega/2$ as can be seen in
Fig.~\ref{fig1}-$b$. On the other hand, calculation of the matrix
elements shows that when the polarization is along $y$, inter-mode
processes are allowed and intra-mode ones are forbidden. Depletions or
gaps appear now at the crossing between Floquet states corresponding to
\textit{different} modes leading to the features observed in
Fig.~\ref{fig2}-$b$ and $c$.

A complementary approach to this problem is possible by using the
\textbf{\textit{k.p}} model, which could be accurate enough for
medium-sized ribbons \cite{Brey}. A careful analysis shows consistent
results: inter-mode processes lead to gaps/depletions located away from
$\pm \hbar \Omega/2$ while polarization along $y$ suppresses the
depletions at $\pm \hbar \Omega/2$. In the bulk limit, as the energy
difference between subbands gets smaller, the crossings between Floquet
states accumulate close to $\pm \hbar \Omega/2$ leading to the same
behavior for both polarizations (along $x$ and $y$). A flavor of this
can be seen in the red dotted lines in Fig.~\ref{fig2}-$b$ and $c$: The
two depletions seen in Fig.~\ref{fig2}-$d$ black line merge when
increasing the laser power.

Another interesting feature is that the metallic modes/subbands in
armchair ribbons are quite insensitive to the radiation (as seen in
Figs.~\ref{fig2} and \ref{fig3}). Hence, very small metallic armchair
ribbons containing only one transport channel within the energy range of
interest will not experience relevant changes in their electronic
properties. We emphasize that this is a peculiar property of armchair
graphene nanoribbons \cite{JPCM2012}.

Figures \ref{fig2} and \ref{fig3} correspond to a laser frequency in the
mid-infrared, which gives an optimum playground to test these
predictions in the laboratory. Going to higher frequencies may help to
achieve device miniaturization but the size of the gaps and depletions
at constant laser power diminishes, whereas at very low frequencies
(THz) the gaps further decrease. Experiments would require temperatures
below $20\unit{K}$ for $P \simeq 1\unit{mW/\mu m^2}$.

%\textit{Conclusions.}
In summary, we show that the interplay between photon-induced inelastic
processes and lateral confinement in graphene nanoribbons leads to
diverse modifications in the band structure and transport properties not
evident in the bulk limit. In the case of moderate sized nanoribbons
(ca. $10 \unit{nm}$), the careful tuning of the polarization direction
may widen the opportunities for achieving control of the electrical
response in optoelectronic devices.

\textit{Acknowledgments.}
We acknowledge support by SeCyT-UNC, ANPCyT-FonCyT (Argentina). LEFFT
acknowledges the support from the Alexander von Humboldt Foundation and
ICTP-Trieste, as well as discussions with G. Usaj.

%\bibliographystyle{prsty}
%\bibliography{jpcm}

\begin{thebibliography}{24}
\expandafter\ifx\csname natexlab\endcsname\relax\def\natexlab#1{#1}\fi
\expandafter\ifx\csname bibnamefont\endcsname\relax
  \def\bibnamefont#1{#1}\fi
\expandafter\ifx\csname bibfnamefont\endcsname\relax
  \def\bibfnamefont#1{#1}\fi
\expandafter\ifx\csname citenamefont\endcsname\relax
  \def\citenamefont#1{#1}\fi
\expandafter\ifx\csname url\endcsname\relax
  \def\url#1{\texttt{#1}}\fi
\expandafter\ifx\csname urlprefix\endcsname\relax\def\urlprefix{URL }\fi
\providecommand{\bibinfo}[2]{#2}
\providecommand{\eprint}[2][]{\url{#2}}

\bibitem{Geim2009} A. K. Geim, Science \textbf{324}, 1934 (2009); A. K. Geim and K. S. Novoselov, Nat. Mat. \textbf{6}, 183 (2007).

\bibitem{CastroNeto2009} A.~H.~Castro Neto, F.~Guinea, N.~M.~R.~Peres, K.~S.
Novoselov and A.~K. Geim, Rev. Mod. Phys. \textbf{81}, 109 (2009).

\bibitem{Dubois2009} S.M.-M. Dubois, Z. Zanolli, X. Declerck, and J.-C. Charlier,  Eur. Phys. J. B \textbf{72}, 1 (2009); J.-C. Charlier, X. Blase, and S. Roche, Rev. Mod. Phys. \textbf{79}, 677 (2007).


\bibitem[{\citenamefont{Bonaccorso et~al.}(2010)\citenamefont{Bonaccorso, Sun,
  Hasan, and Ferrari}}]{Bonaccorso2010}
\bibinfo{author}{\bibfnamefont{F.}~\bibnamefont{Bonaccorso}},
  \bibinfo{author}{\bibfnamefont{Z.}~\bibnamefont{Sun}},
  \bibinfo{author}{\bibfnamefont{T.}~\bibnamefont{Hasan}}, \bibnamefont{and}
  \bibinfo{author}{\bibfnamefont{A.~C.} \bibnamefont{Ferrari}},
  \bibinfo{journal}{Nat Photon} \textbf{\bibinfo{volume}{4}},
  \bibinfo{pages}{611} (\bibinfo{year}{2010}).% ISSN \bibinfo{issn}{1749-4885},
%  \urlprefix\url{http://dx.doi.org/10.1038/nphoton.2010.186}.

\bibitem[{\citenamefont{Xia et~al.}(2009)\citenamefont{Xia, Mueller, Lin,
  Valdes-Garcia, and Avouris}}]{Xia2009}
\bibinfo{author}{\bibfnamefont{F.}~\bibnamefont{Xia}},
  \bibinfo{author}{\bibfnamefont{T.}~\bibnamefont{Mueller}},
  \bibinfo{author}{\bibfnamefont{Y.-m.} \bibnamefont{Lin}},
  \bibinfo{author}{\bibfnamefont{A.}~\bibnamefont{Valdes-Garcia}},
  \bibnamefont{and} \bibinfo{author}{\bibfnamefont{P.}~\bibnamefont{Avouris}},
  \bibinfo{journal}{Nat Nano} \textbf{\bibinfo{volume}{4}},
  \bibinfo{pages}{839} (\bibinfo{year}{2009}).%, ISSN \bibinfo{issn}{1748-3387}.
%  \urlprefix\url{http://dx.doi.org/10.1038/nnano.2009.292}.

\bibitem[{\citenamefont{Gabor et~al.}(2011)\citenamefont{Gabor, Song, Ma, Nair,
  Taychatanapat, Watanabe, Taniguchi, Levitov, and
  Jarillo-Herrero}}]{Gabor2011}
\bibinfo{author}{\bibfnamefont{N.~M.} \bibnamefont{Gabor}},
  \bibinfo{author}{\bibfnamefont{J.~C.~W.} \bibnamefont{Song}},
  \bibinfo{author}{\bibfnamefont{Q.}~\bibnamefont{Ma}},
  \bibinfo{author}{\bibfnamefont{N.~L.} \bibnamefont{Nair}},
  \bibinfo{author}{\bibfnamefont{T.}~\bibnamefont{Taychatanapat}},
  \bibinfo{author}{\bibfnamefont{K.}~\bibnamefont{Watanabe}},
  \bibinfo{author}{\bibfnamefont{T.}~\bibnamefont{Taniguchi}},
  \bibinfo{author}{\bibfnamefont{L.~S.} \bibnamefont{Levitov}},
  \bibnamefont{and}
  \bibinfo{author}{\bibfnamefont{P.}~\bibnamefont{Jarillo-Herrero}},
  \bibinfo{journal}{Science} \textbf{\bibinfo{volume}{334}},
  \bibinfo{pages}{648} (\bibinfo{year}{2011}).
%  \urlprefix\url{http://www.sciencemag.org/content/334/6056/648.abstract}.

\bibitem[{\citenamefont{Karch et~al.}(2011)\citenamefont{Karch, Drexler,
  Olbrich, Fehrenbacher, Hirmer, Glazov, Tarasenko, Ivchenko, Birkner, Eroms
  et~al.}}]{Karch2011}
\bibinfo{author}{\bibfnamefont{J.}~\bibnamefont{Karch}},
  \bibinfo{author}{\bibfnamefont{C.}~\bibnamefont{Drexler}},
  \bibinfo{author}{\bibfnamefont{P.}~\bibnamefont{Olbrich}},
  \bibinfo{author}{\bibfnamefont{M.}~\bibnamefont{Fehrenbacher}},
  \bibinfo{author}{\bibfnamefont{M.}~\bibnamefont{Hirmer}},
  \bibinfo{author}{\bibfnamefont{M.~M.} \bibnamefont{Glazov}},
  \bibinfo{author}{\bibfnamefont{S.~A.} \bibnamefont{Tarasenko}},
  \bibinfo{author}{\bibfnamefont{E.~L.} \bibnamefont{Ivchenko}},
  \bibinfo{author}{\bibfnamefont{B.}~\bibnamefont{Birkner}},
  \bibinfo{author}{\bibfnamefont{J.}~\bibnamefont{Eroms}},
  \bibnamefont{et~al.}, \bibinfo{journal}{Phys. Rev. Lett.}
  \textbf{\bibinfo{volume}{107}}, \bibinfo{pages}{276601}
  (\bibinfo{year}{2011}).
%  \urlprefix\url{http://link.aps.org/doi/10.1103/PhysRevLett.107.276601}.

\bibitem[{\citenamefont{Koppens et~al.}(2011)\citenamefont{Koppens, Chang, and
  García~de Abajo}}]{Koppens2011}
\bibinfo{author}{\bibfnamefont{F.~H.~L.} \bibnamefont{Koppens}},
  \bibinfo{author}{\bibfnamefont{D.~E.} \bibnamefont{Chang}}, \bibnamefont{and}
  \bibinfo{author}{\bibfnamefont{F.~J.} \bibnamefont{García~de Abajo}},
  \bibinfo{journal}{Nano Lett.} \textbf{\bibinfo{volume}{11}},
  \bibinfo{pages}{3370} (\bibinfo{year}{2011}). %, ISSN \bibinfo{issn}{1530-6984}.
%  \urlprefix\url{http://dx.doi.org/10.1021/nl201771h}.

\bibitem{Ren2009} L. Ren, C. L. Pint, L. G. Booshehri, W. D. Rice, X. Wang, D. J. Hilton, K. Takeya, I. Kawayama, M. Tonouchi, R. H. Hauge and %J. Kono, Nano Letters \textbf{9}, 2610 (2009). 

\bibitem{McIver2012} J. W. McIver, D. Hsieh, H. Steinberg, P. Jarillo-Herrero and N. Gedik, Nat. Nanotech. \textbf{7}, 96 (2012).

\bibitem{Chen2012} J. Chen,	M. Badioli, P. Alonso-Gonzalez, S. Thongrattanasiri, F. Huth, J. Osmond, M. Spasenovic, A. Centeno, A. Pesquera, Ph. Godignon, A. Zurutuza Elorza, N. Camara, F. J. Garcia de Abajo, R. Hillenbrand and F. H. L. Koppens, Nature (2012), doi:10.1038/nature11254.

\bibitem[{\citenamefont{Syzranov et~al.}(2008)\citenamefont{Syzranov, Fistul,
  and Efetov}}]{Syzranov2008}
\bibinfo{author}{\bibfnamefont{S.~V.} \bibnamefont{Syzranov}},
  \bibinfo{author}{\bibfnamefont{M.~V.} \bibnamefont{Fistul}},
  \bibnamefont{and} \bibinfo{author}{\bibfnamefont{K.~B.}
  \bibnamefont{Efetov}}, \bibinfo{journal}{Phys. Rev. B}
  \textbf{\bibinfo{volume}{78}}, \bibinfo{pages}{045407}
  (\bibinfo{year}{2008}).
%  \urlprefix\url{http://link.aps.org/doi/10.1103/PhysRevB.78.045407}.

\bibitem[{\citenamefont{Oka and Aoki}(2009)}]{Oka2009}
\bibinfo{author}{\bibfnamefont{T.}~\bibnamefont{Oka}} \bibnamefont{and}
  \bibinfo{author}{\bibfnamefont{H.}~\bibnamefont{Aoki}},
  \bibinfo{journal}{Phys. Rev. B} \textbf{\bibinfo{volume}{79}},
  \bibinfo{pages}{081406} (\bibinfo{year}{2009}).
%  \urlprefix\url{http://link.aps.org/doi/10.1103/PhysRevB.79.081406}.

\bibitem[{\citenamefont{Calvo et~al.}(2011)\citenamefont{Calvo, Pastawski,
  Roche, and Torres}}]{Calvo2011}
\bibinfo{author}{\bibfnamefont{H.~L.} \bibnamefont{Calvo}},
  \bibinfo{author}{\bibfnamefont{H.~M.} \bibnamefont{Pastawski}},
  \bibinfo{author}{\bibfnamefont{S.}~\bibnamefont{Roche}}, \bibnamefont{and}
  \bibinfo{author}{\bibfnamefont{L.~E. F.} \bibnamefont{Foa Torres}},
  \bibinfo{journal}{Appl. Phys. Lett.} \textbf{\bibinfo{volume}{98}},
  \bibinfo{pages}{232103} (\bibinfo{year}{2011}).
%  \urlprefix\url{http://dx.doi.org/10.1063/1.3597412}.

\bibitem[{\citenamefont{Zhou and Wu}(2011)}]{Zhou2011}
\bibinfo{author}{\bibfnamefont{Y.}~\bibnamefont{Zhou}} \bibnamefont{and}
  \bibinfo{author}{\bibfnamefont{M.~W.} \bibnamefont{Wu}},
  \bibinfo{journal}{Phys. Rev. B} \textbf{\bibinfo{volume}{83}},
  \bibinfo{pages}{245436} (\bibinfo{year}{2011}).
%  \urlprefix\url{http://link.aps.org/doi/10.1103/PhysRevB.83.245436}.

\bibitem[{\citenamefont{Busl et~al.}(2012)\citenamefont{Busl, Platero, and
  Jauho}}]{Busl2012}
\bibinfo{author}{\bibfnamefont{M.}~\bibnamefont{Busl}},
  \bibinfo{author}{\bibfnamefont{G.}~\bibnamefont{Platero}}, \bibnamefont{and}
  \bibinfo{author}{\bibfnamefont{A.-P.} \bibnamefont{Jauho}},
  \bibinfo{journal}{Phys. Rev. B} \textbf{\bibinfo{volume}{85}},
  \bibinfo{pages}{155449} (\bibinfo{year}{2012}).
%  \urlprefix\url{http://link.aps.org/doi/10.1103/PhysRevB.85.155449}.

\bibitem[{\citenamefont{Kitagawa et~al.}(2011)\citenamefont{Kitagawa, Oka,
  Brataas, Fu, and Demler}}]{Kitagawa2011}
\bibinfo{author}{\bibfnamefont{T.}~\bibnamefont{Kitagawa}},
  \bibinfo{author}{\bibfnamefont{T.}~\bibnamefont{Oka}},
  \bibinfo{author}{\bibfnamefont{A.}~\bibnamefont{Brataas}},
  \bibinfo{author}{\bibfnamefont{L.}~\bibnamefont{Fu}}, \bibnamefont{and}
  \bibinfo{author}{\bibfnamefont{E.}~\bibnamefont{Demler}},
  \bibinfo{journal}{Phys. Rev. B} \textbf{\bibinfo{volume}{84}},
  \bibinfo{pages}{235108} (\bibinfo{year}{2011}).
%  \urlprefix\url{http://link.aps.org/doi/10.1103/PhysRevB.84.235108}.

\bibitem[{\citenamefont{Gu et~al.}(2011)\citenamefont{Gu, Fertig, Arovas, and
  Auerbach}}]{Gu2011}
\bibinfo{author}{\bibfnamefont{Z.}~\bibnamefont{Gu}},
  \bibinfo{author}{\bibfnamefont{H.~A.} \bibnamefont{Fertig}},
  \bibinfo{author}{\bibfnamefont{D.~P.} \bibnamefont{Arovas}},
  \bibnamefont{and} \bibinfo{author}{\bibfnamefont{A.}~\bibnamefont{Auerbach}},
  \bibinfo{journal}{Phys. Rev. Lett.} \textbf{\bibinfo{volume}{107}},
  \bibinfo{pages}{216601} (\bibinfo{year}{2011}).
%  \urlprefix\url{http://link.aps.org/doi/10.1103/PhysRevLett.107.216601}.

\bibitem[{\citenamefont{Dóra et~al.}(2012)\citenamefont{Dóra, Cayssol, Simon,
  and Moessner}}]{Dora2012}
\bibinfo{author}{\bibfnamefont{B.}~\bibnamefont{Dóra}},
  \bibinfo{author}{\bibfnamefont{J.}~\bibnamefont{Cayssol}},
  \bibinfo{author}{\bibfnamefont{F.}~\bibnamefont{Simon}}, \bibnamefont{and}
  \bibinfo{author}{\bibfnamefont{R.}~\bibnamefont{Moessner}},
  \bibinfo{journal}{Phys. Rev. Lett.} \textbf{\bibinfo{volume}{108}},
  \bibinfo{pages}{056602} (\bibinfo{year}{2012}).
%  \urlprefix\url{http://link.aps.org/doi/10.1103/PhysRevLett.108.056602}.

\bibitem[{\citenamefont{Suarez~Morell and Foa~Torres}(2012)}]{SuarezMorell2012}
\bibinfo{author}{\bibfnamefont{E.}~\bibnamefont{Suarez~Morell}}
  \bibnamefont{and} \bibinfo{author}{\bibfnamefont{L.~E.~F.}
  \bibnamefont{Foa~Torres}}, \bibinfo{journal}{Physical Review B}
  \textbf{\bibinfo{volume}{86}},
  \bibinfo{pages}{125449} (\bibinfo{year}{2012}).

\bibitem[{\citenamefont{Abergel and Chakraborty}(2009)}]{Abergel2009}
\bibinfo{author}{\bibfnamefont{D.~S.~L.} \bibnamefont{Abergel}}
  \bibnamefont{and}
  \bibinfo{author}{\bibfnamefont{T.}~\bibnamefont{Chakraborty}},
  \bibinfo{journal}{Appl. Phys. Lett.} \textbf{\bibinfo{volume}{95}},
  \bibinfo{pages}{062107} (\bibinfo{year}{2009}).
%  \urlprefix\url{http://dx.doi.org/10.1063/1.3205117}.

\bibitem[{\citenamefont{Kibis}(2010)}]{Kibis2010}
\bibinfo{author}{\bibfnamefont{O.~V.} \bibnamefont{Kibis}},
  \bibinfo{journal}{Phys. Rev. B} \textbf{\bibinfo{volume}{81}},
  \bibinfo{pages}{165433} (\bibinfo{year}{2010}).
%  \urlprefix\url{http://link.aps.org/doi/10.1103/PhysRevB.81.165433}.

\bibitem[{\citenamefont{Savel’ev and Alexandrov}(2011)}]{Savelev2011}
\bibinfo{author}{\bibfnamefont{S.~E.} \bibnamefont{Savel’ev}}
  \bibnamefont{and} \bibinfo{author}{\bibfnamefont{A.~S.}
  \bibnamefont{Alexandrov}}, \bibinfo{journal}{Phys. Rev. B}
  \textbf{\bibinfo{volume}{84}}, \bibinfo{pages}{035428}
  (\bibinfo{year}{2011}).
%  \urlprefix\url{http://link.aps.org/doi/10.1103/PhysRevB.84.035428}.

\bibitem[{\citenamefont{Iurov et~al.}(2012)\citenamefont{Iurov, Gumbs, Roslyak,
  and Huang}}]{Iurov2012}
\bibinfo{author}{\bibfnamefont{A.}~\bibnamefont{Iurov}},
  \bibinfo{author}{\bibfnamefont{G.}~\bibnamefont{Gumbs}},
  \bibinfo{author}{\bibfnamefont{O.}~\bibnamefont{Roslyak}}, \bibnamefont{and}
  \bibinfo{author}{\bibfnamefont{D.}~\bibnamefont{Huang}},
  \bibinfo{journal}{Journal of Physics: Condensed Matter}
  \textbf{\bibinfo{volume}{24}}, \bibinfo{pages}{015303}
  (\bibinfo{year}{2012}). % ISSN \bibinfo{issn}{0953-8984}.
%  \urlprefix\url{http://stacks.iop.org/0953-8984/24/i=1/a=015303}.

\bibitem[{\citenamefont{Liu et~al.}(2012)\citenamefont{Liu, Su, Wang, and
  Deng}}]{Liu2012}
\bibinfo{author}{\bibfnamefont{J.-T.} \bibnamefont{Liu}},
  \bibinfo{author}{\bibfnamefont{F.-H.} \bibnamefont{Su}},
  \bibinfo{author}{\bibfnamefont{H.}~\bibnamefont{Wang}}, \bibnamefont{and}
  \bibinfo{author}{\bibfnamefont{X.-H.} \bibnamefont{Deng}},
  \bibinfo{journal}{New Journal of Physics} \textbf{\bibinfo{volume}{14}},
  \bibinfo{pages}{013012} (\bibinfo{year}{2012}). %, ISSN
%  \bibinfo{issn}{1367-2630}.
%  \urlprefix\url{http://stacks.iop.org/1367-2630/14/i=1/a=013012}.

\bibitem[{\citenamefont{San-Jose et~al.}()\citenamefont{San-Jose, Prada,
  Schomerus, and Kohler}}]{San-Jose2012}
\bibinfo{author}{\bibfnamefont{P.}~\bibnamefont{San-Jose}},
  \bibinfo{author}{\bibfnamefont{E.}~\bibnamefont{Prada}},
  \bibinfo{author}{\bibfnamefont{H.}~\bibnamefont{Schomerus}},
  \bibnamefont{and} \bibinfo{author}{\bibfnamefont{S.}~\bibnamefont{Kohler}},
  \bibinfo{journal}{Appl. Phys. Lett.} \textbf{\bibinfo{volume}{101}},
  \bibinfo{pages}{153506} (\bibinfo{year}{2012}).

\bibitem{SaitoBook}  R. Saito,  G. Dresselhaus, and M. S. Dresselhaus, 1998, Physical Properties of Carbon Nanotubes (Imperial College Press, London).

\bibitem[{\citenamefont{Kohler et~al.}(2005)\citenamefont{Kohler, Lehmann, and
  Hänggi}}]{Kohler2005}
\bibinfo{author}{\bibfnamefont{S.}~\bibnamefont{Kohler}},
  \bibinfo{author}{\bibfnamefont{J.}~\bibnamefont{Lehmann}}, \bibnamefont{and}
  \bibinfo{author}{\bibfnamefont{P.}~\bibnamefont{Hänggi}},
  \bibinfo{journal}{Physics Reports} \textbf{\bibinfo{volume}{406}},
  \bibinfo{pages}{379} (\bibinfo{year}{2005}).%, ISSN \bibinfo{issn}{0370-1573}.
  %\urlprefix\url{http://www.sciencedirect.com/science/article/pii/S03701573040%
%05071}.

\bibitem[{\citenamefont{Foa~Torres}(2005)}]{FoaTorres2005}
\bibinfo{author}{\bibfnamefont{L.~E.~F.} \bibnamefont{Foa~Torres}},
  \bibinfo{journal}{Phys. Rev. B} \textbf{\bibinfo{volume}{72}},
  \bibinfo{pages}{245339} (\bibinfo{year}{2005}).
%  \urlprefix\url{http://link.aps.org/doi/10.1103/PhysRevB.72.245339}.

\bibitem{JPCM2012} H. L. Calvo, P. Perez Piskunow, H. M. Pastawski, S. Roche, L. E. F. Foa Torres, unpublished.

\bibitem{Shirley} J. H. Shirley, Phys. Rev. \textbf{138}, B979 (1965); H. Sambe, Phys. Rev. A \textbf{7}, 2203 (1973).

\bibitem{Brey} L. Brey, H.A. Fertig, Phys. Rev. B {\bf 73}, 235411 (2006); P. Marconcini and P. Maccuci, Riv. del Nuovo Cimento {\bf 34}, 489 (2011).

\bibitem[{\citenamefont{Rocha et~al.}(2010)\citenamefont{Rocha, Torres, and
  Cuniberti}}]{Rocha2010}
\bibinfo{author}{\bibfnamefont{C.~G.} \bibnamefont{Rocha}},
  \bibinfo{author}{\bibfnamefont{L.~E. F.} \bibnamefont{Foa Torres}},
  \bibnamefont{and}
  \bibinfo{author}{\bibfnamefont{G.}~\bibnamefont{Cuniberti}},
  \bibinfo{journal}{Phys. Rev. B} \textbf{\bibinfo{volume}{81}},
  \bibinfo{pages}{115435} (\bibinfo{year}{2010}).
%  \urlprefix\url{http://link.aps.org/doi/10.1103/PhysRevB.81.115435}.

\bibitem[{\citenamefont{Zhao and Guo}(2009)}]{Zhao2009}
\bibinfo{author}{\bibfnamefont{P.}~\bibnamefont{Zhao}} \bibnamefont{and}
  \bibinfo{author}{\bibfnamefont{J.}~\bibnamefont{Guo}},
  \bibinfo{journal}{Journal of Applied Physics} \textbf{\bibinfo{volume}{105}},
  \bibinfo{eid}{034503} (\bibinfo{year}{2009}).
%  \urlprefix\url{http://link.aip.org/link/?JAP/105/034503/1}.


\end{thebibliography}

\end{document}